# A Search for Pulsars between 0.3 and 8 MeV using INTEGRAL

Joseph F. Dolan

Department of Astronomy, San Diego State University

San Diego, CA 92182-1221

e-mail: tejfd@sciences.sdsu.edu

# Abstract

The spectra of only a few pulsars are known in the low-energy gamma-ray region around 1 Mev. We searched for point sources of gamma-radiation in the 0.3–1 and 1–8 Mev bandpasses at the location of known radio and X-ray pulsars. Single channel photometry using the SPI spectrometer on *Integral*, obtained using a coded-aperture mask, is feasible if calibrated using known fluxes from point sources. The calibration was carried out using the Crab nebula and verified using the Vela pulsar. The locations of a representative sample of pulsar types were then observed. Other than the Crab and Vela pulsars, no sources were detected. Upper limits are given on the time-averaged flux from the location of 15 other pulsars; most lie below previously published upper limits. Six of the pulsars for which we determine upper limits at 1 MeV have been detected above 100 MeV by *Fermi*. The upper limits we report here help to describe the spectrum of these pulsars at all wavelengths.

# 1. Introduction

Determining the structure of neutron stars is one path to understanding the physics of matter at super-nuclear densities (Stock 1989). Pulsars are an important class of neutron stars whose spectra have been detected at frequencies ranging from radio to gamma-ray. They are "the second most numerous class of objects detected and identified in high-energy gamma rays" (Brazier et al. 1994). Knowing the spectrum of pulsars in every wavelength region in which they radiate is a necessary step towards achieving an understanding of their emission mechanism and the underlying particle acceleration process. A basic issue is whether the high-energy emission arises near the surface of the neutron star (e.g., Daugherty 1996) or at a significant fraction of the distance to the velocity of light cylinder (e.g. Smith 1986; Romani 1996).

Only the Crab pulsar (B0531+21), the Vela pulsar (B0833-45), and PSR B1509-58 have been detected in the low-energy gamma-ray region around 1 MeV (Strong et al. 1993; Graser & Schonfelder 1982; Tumer et al. 1984; Matz et al. 1994). Upper limits exist on the flux of several other pulsars in this energy region (Carriminana et al. 1995). An observing program using the *Fermi* satellite (Abdo et al. 2009a; 2009b; 2010a; 2010b) attempted to detect pulsars belonging to several different sub-classes above 10 MeV. A survey to detect, or place upper limits on, pulsar fluxes in the 1 - 8 Mev region was carried out as a pathfinder for target selection for this program. We used archival data from the *Integral* satellite to observe the location of known radio and X-ray pulsars in the low-energy gamma-ray region. Six of the pulsars with upper limits we report here at 1 MeV have been detected above 100 MeV by *Fermi* (Abdo et al. 2009a; 2010a).

## 1.1. Photometry using SPI

The Spectrometer on *Integral* (SPI) uses cooled Ge detectors, a coded-mask aperture, and an active anti-coincidence shield to detect photons from 20 keV to 8 Mev (Vedrenne et al. 2003). The angular resolution of the instrument for point sources is 2.5 deg FWHM. A description of the instrument's performance in orbit is given by Roques et al. (2003).

The location of point sources in the field of view of SPI and their associated fluxes as a function of energy are derived in standard analysis routines available from the *Integral* project using a maximum likelihood algorithm (Skinner & Connell 2003; Strong 2003).We investigated the location of known pulsars in two energy bandpasses, 0.3 – 1 and 1 – 8 MeV, using SPI as a single-channel photometer in each. It is well known, however, that a maximum likelihood fit to the intensity of a point source in a single bandpass using SPI (i.e. photometry) produces a biased flux estimate (Carson 2006; Smith 2006). The bias is consistent, however, in the sense that the flux value derived is a constant fraction of incident flux from the source. Hence, gamma-ray photometry can be done using SPI in a single bandpass by calibrating the bias with a source having a known flux in that bandpass.

We calibrated SPI in the 03 – 1 and 1 – 8 MeV bandpasses using observations of the total flux (time average of the pulsed plus dc components) from the Crab nebula reported by Strong et al. (1993) and Graser & Schonfelder (1982). The *Integral* observations of the Crab we used totaled 1 x $10^6$ s live time and were obtained on 2003 February 16-27, 2004 March 5-7, and September 27-30; and 2005 March 29-30. The ratio of the incident flux reported in the literature to the counting rate (in $cm^{-2}$ $s^{-1}$) returned by the analysis programs for the source was taken as the conversion factor between the two for each *Integral* bandpass. The calibration was then verified by deriving the flux of the Vela pulsar in each bandpass and comparing it with that reported by Tumer et al. (1984). The Vela observations we used totaled 1.1 x $10^6$ s live time and were obtained on 2003 June 12-21 and July 1-6. The fluxes we derived for the Vela pulsr in both bandpasses were consistent with those reported by Tume et al. The fluxes we observed from the Crab nebula and Vela pulsar are given in Table 1.

## 1.2. Target selection

We observed a representative sample of several different classes of known radio and X-ray pulsars. Parameters defing pulsar characteristics can be found in the Australian National Telescope Facility Pulsar Catalog (2006). Our sample included

- pulsars previously detected in the high-energy gamma-ray region (Thompson et al. 1994): Crab (B0535+21); Geminga (J0633+1746); Vela (B0833-45); B1055-52; B1509-58; B1706-44

- pulsars with a large spin-down rate ($d\nu/dt$) or energy loss rate ($dE/dt = [4 \pi^2 I\, dp/dt] / p^3$, where I is the moment of inertia of the neutron star): J0205+6449; J0537-6910; B0540-69; J1833-1034

- pulsars in a binary system: B1259-63 (Manchester et al. 1995); B1913+16 (Hulse & Taylor 1974)

- pulsars with a millisecond period: I00291+5934 (Falanga et al. 2005)

- nearby pulsars (with small values of dispersion measure in the radio region): B1929+10

- pulsars with a small characteristic age ($\tau = [p/2]\, dp/dt$): B1727-47

- pulsars detected in hard X-rays: J0142+61 (Kuiper et al. 2006).

## 2. Results

No sources were detected at the location of the 15 pulsars observed in addition to the Crab and Vela. Upper limits on the flux density of each are given in Table 1. The 2σ upper limits are given in nJy. [1 nJy = $10^{-32}$ erg cm$^{-2}$ s$^{-1}$ Hz$^{-1}$.] For reference, 1 nJy corresponds to 2.1 x $10^{-5}$ (3.5 x $10^{-6}$) photons cm$^{-2}$ s$^{-1}$ in the 0.3 – 1 (1 – 8) Mev bandpass.

Table 1.

Time-averaged flux density of pulsars observed with *Integral.**

| Pulsar | $S_\nu$ (nJy) | |
|---|---|---|
| | 0.3-1 MeV | 1-8 MeV |
| Crab | 7310 ± 240 | 2810 ± 110 |
| Vela | 142 ± 39 | 28 ± 18 |
| Geminga | <195 | <90 |
| I00291+5934 | <160 | <75 |
| 4U0142+61 | <220 | <130 |
| J0205+6449 | <300 | <130 |
| J0537-6910 | <120 | <90 |
| B0540-69 | <135 | <100 |
| B1055-52 | <65 | <35 |
| B1259-63 | <75 | <40 |
| B1509-58 | <80 | <40 |
| B1706-44 | <65 | <30 |
| B1727-47 | <65 | <30 |
| J1833-103 | <200 | <85 |
| B1913+16 | <80 | <55 |
| B1929+19 | <120 | <65 |
| B1951+32 | <75 | <40 |

* Non-detected sources are given as 2σ upper limits

## 3. Discussion

Most of the 2σ upper limits in Table 1 lie below those previously published in the literature. The upper limits we report around 1 Mev on the six pulsars detected by *Fermi* above 100 Mev complete the description of these pulsars' spectrum at all wavelengths.

<u>I0029+5934</u>: a transient X-ray source discovered by Shaw et al. (2005) in observations made with the *Integral* satellite. It is classified as an accreting pulsar with a 1.67 ms pulse period (Galloway et al. 2005). The upper limits in Table 1 refer to the epoch 2004 December 4.66 – 10.90, when the source was in outburst in the 20 – 100 keV bandpass (Falanga et al. 2005).

<u>J0205+6449</u>: detected above 100 MeV by *Fermi* (Abdo et al. 2010a).

<u>B1055-52</u>: Carraminana et al. (1995) give an upper limit of 10 nJy between 1 and 3 MeV using *Comptel* observations. Detected above 100 MeV by *Fermi* (Abdo et al. 2009a; 2010a).

<u>B1259-63</u>: Carraminana et al. (1995) given an upper limit of 105 nJy (1-3 MeV); our upper limit lies below this value.

<u>B1509-58</u>: a pulsed flux from B1509-58 at the period of the radio pulsar was detected by Matz et al. (1994) using *Osse*. The phase-averaged flux density was 40 ± 29 nJy (0.35-0.6 MeV) ;41 ± 29 nJy (0.7-1.2 MeV); and 37 ± 23 nJy (1.2-5 MeV). A pulsed source was also detected by Kuiper et al. (1999) using *Comptel*;

they give the flux in the "pulsed" interval extending over half the pulse period of 37 ± 7 nJy (0.75-3MeV) and 15 ± 3 nJy (3-10 MeV). The minimum time-averaged (dc) flux would presumably be half of these values if no detectable flux were emitted by the pulsar outside of the "pulsed" interval. Our upper limits on the time-averaged flux are consistent with these results. 1509-58 was detected above 100 MeV by Fermi (Abdo et al. 2009a; 2010a).

B1706-44: Carrimanana et al. (1995) five an upper limit of 160 nJy (1-3 Mev); our upper limit lies below this value. Detected above 100 MeV by *Fermi* (Abdo et al. 2009a; 2010a).

J1833-1034: detected above 100 MeV by *Fermi* (Abdo et al. 2010a).

B1929+10: our 1-3 MeV upper limit lies below the 130 nJy UL given by Carrimanana et al. (1995).

B1951+32: detected above 100 MeV by *Fermi* (Abdo et al. 2009a; 2010a).

**Acknowledgements.** This paper is based on observations taken from the *Integral* archive. Part of this work was carried out when the author was at NASA's Goddard Space Flight Center. We thank V. Beckman and S. Sturner for their assistance in using the archive, and A. K. Harding for a discussion regarding B1509-58.